\documentclass[oldversion]{aa}  
\usepackage{graphicx}
\usepackage{txfonts}
\usepackage{natbib}
\begin{document}
\title{Improved orbital solution and masses for the very low-mass multiple system LHS~1070\thanks{Based on observations made with ESO Telescopes at the Paranal Observatory under programme ID 60.A-9078(A) and 79.C-0106(A), as well as data collected at Subaru Telescope, which is operated by the National Astronomical Observatory of Japan. Based also on observations obtained at the Canada-France-Hawaii Telescope (CFHT) which is operated by the National Research Council of Canada, the Institut National des Sciences de l’Univers of the Centre National de la Recherche Scientiﬁque of France, and the University of
Hawaii.
}}
\titlerunning{Improved orbital solution and masses for LHS~1070}

   \author{A. Seifahrt\inst{1,2}
           \and
           T. R\"{o}ll\inst{2}
           \and
           R. Neuh\"{a}user\inst{2}
           \and
           A. Reiners\inst{1}
          \and
           F. Kerber\inst{3}
          \and
           H. U. K\"{a}ufl\inst{3}
          \and 
           R. Siebenmorgen\inst{3}
          \and
           A. Smette\inst{4}}


   \institute{Universit\"at G\"ottingen, Institut f\"ur Astrophysik, Friedrich-Hund-Platz 1, D-37077 G\"ottingen, Germany\\
     \email{seifahrt@astro.physik.uni-goettingen.de}
     \and
     Universit\"at Jena, Astrophysikalisches Institut und Universit\"ats-Sternwarte, Schillerg\"asschen 2, D-07745 Jena, Germany
     \and
     ESO, Karl-Schwarzschild-Str. 2, D-85748 Garching, Germany
     \and
     ESO, Alonso de C\'ordova 3107, Vitacura, Casilla 19001, Santiago 19, Chile
   }
   
   \date{as of \today}
   
 
  \abstract
  { We present a refined orbital solution for the components A, B, and
    C of the nearby late-M type multiple system LHS~1070. By combining 
    astrometric datapoints from NACO/VLT, CIAO/SUBARU, and PUEO/CFHT,
    as well as a radial velocity measurement from the newly commissioned near infrared 
    high-resolution spectrograph CRIRES/VLT, we achieve a very precise orbital 
    solution for the B and C components and a first realistic constraint 
    on the much longer orbit of the A-BC system. Both orbits appear to be co-planar. 
    Masses for the B and C components calculated from the new orbital 
    solution ($M_{B+C} = 0.157 \pm 0.009~\mathrm{M_{\sun}}$) are in excellent 
    agreement with theoretical models, but do not match empirical mass-luminosity tracks. 
    The preliminary orbit of the A-BC system reveals no mass excess 
    for the A component, giving no indication for a previously proposed fourth 
    (D) component in LHS~1070.}


   \maketitle
%

\section{Introduction}
LHS 1070 (GJ 2005, LP 881-64) is a low-mass triple system at a distance of 
7.72 $\pm$ 0.15 pc from the Sun \citep{Costa05}. It consists of two components 
with masses close to the substellar limit (LHS 1070 B and C) with spectral types
of M8.5 to M9, orbited by a slightly more massive M5.5 type component, LHS 1070 A.

This system is important when defining the empirical mass--luminosity
relationship at the low-mass end of the main sequence. Few reliable and accurate mass determinations,
such as dynamical masses from binary and multiple systems, exist below 0.1 M$_{\sun}$, which makes
LHS 1070 an important anchor point, especially given its age of several Gigayears and its
proximity to the Sun.  

\citet{Leinert01} have presented the first orbital solution for the LHS 1070 BC system based on
speckle measurements and early adaptive optics images. Deriving a combined mass of 0.138 
M$_{\sun}$ for LHS 1070 BC, they put the system in the context of the empirical and theoretical 
mass--luminosity relationships presented in \citet{Leinert00}. Given the much longer
orbital period of the outer, more massive component LHS 1070 A, mass estimates for LHS 1070 A
and a first reliable solution for the outer orbit were beyond reach in 2001. Nevertheless,
\citet{Leinert01} succeeded in giving rough constraints on the outer orbit, suggesting 
co-planarity with the inner orbit of LHS 1070 BC.

Late-type double and multiple systems are also important probes for the connection of rotation
and magnetic activity at the transition from partly to fully convective stars. During the science 
verification phase of CRIRES at the VLT \citep{CRIRES}, we obtained spatially resolved measurements of the
rotational velocity and magnetic activity for all three components of LHS 1070 \citep[see][for details]{Reiners07c}. 
This measurement is essentially free of the uncertainties induced from the otherwise unknown 
inclination angle when comparing rotation and magnetic activity for objects of different spectral 
types, since the rotation axes of the components in this triple system are most likely spatially 
well-aligned. The CRIRES/VLT datapoint also adds important information on the orbit of the triple 
system, removing the ambiguity in the true spatial orientation. 

In this paper we present our results from an improved orbital fit of LHS 1070 BC and first 
constraints on the orbit of LHS 1070 A around the barycentre of the BC subsystem. 
Our observations, as well as extensive archival data, are presented in Section 2. The orbital solutions 
and masses derived from these data are discussed in Section 3. A summary of the most important 
results is given in Section 4.
 
\section{Observations and data reduction}
\begin{table*}[ht!]
\caption{\label{tab:observations} {Observation log and obtained relative positions for LHS1070 C ($X_C,Y_C$) with respect to component B and for the barycentre of LHS 1070 BC ($X_{BC},Y_{BC}$) with respect to component A.}}
\begin{center}

\begin{minipage}[t]{1.95\columnwidth}
\renewcommand{\footnoterule}{}  
\begin{tabular}{llllllrrrr}
    \hline
    \hline
    \noalign{\smallskip}
    Date & Instrument & Pixel scale & Filter  & Prog.ID. & P.I. & $X_C$\footnote{$X_C$ and $Y_C$ are accurate to $\pm$1.5~mas}\,\, & $Y_C$\,\,\,& $X_{BC}$\footnote{$X_{BC}$ and $Y_{BC}$ are accurate to $\pm$6~mas}\,\, & $Y_{BC}$\,\,\,\\
         &            & (mas/pixel) &         &          &      &     (mas)   & (mas)      &     (mas)   & (mas)  \\
    \noalign{\smallskip}
    \hline
    \noalign{\smallskip}
\multicolumn{10}{c}{Observations retrieved from archives}\\
\noalign{\smallskip}
2000-08-16 & PUEO/CFHT & 34.9  & K$^\prime$    & 00AD99 & Forveille    &   25.0 & 229.1 & 1520 & 206 \\
2000-12-12 & PUEO/CFHT & 34.9  & K             & 00BH12 & Roddier      &  -24.6 & 218.5 & 1526 & 224 \\
2002-07-24 & PUEO/CFHT & 34.9  & K$^\prime$    & 02AF43 & Beuzit       & -265.2 & 126.7 & 1541 & 315 \\
2002-09-11 & PUEO/CFHT & 34.9  & H, K$^\prime$ & 02BF27 & Beuzit       & -281.8 & 116.9 & 1535 & 326 \\
2002-10-21 & NACO/VLT  & 13.24 & J, Ks, NB2.17 & 70.C-0476(A) & Bailer-Jones & -295.8 & 109.3 & 1537 & 334 \\
2002-11-15 & NACO/VLT  & 13.24 & NB2.17        & 70.C-0476(A) & Bailer-Jones & -303.3 & 103.9 & 1540 & 337 \\
2002-12-16 & NACO/VLT  & 13.24 & J, Ks     & 70.C-0704(A) & Leinert      & -314.3 &  97.3 & 1538 & 342 \\
2003-01-25 & NACO/VLT  & 13.24 & J, Ks, NB2.17 & 70.C-0476(A) & Bailer-Jones & -327.1 &  89.2 & 1536 & 344 \\
2003-05-24 & NACO/VLT  & 13.24 & J, Ks, NB2.17 & 70.C-0476(A) & Bailer-Jones & -362.4 &  59.0 & 1534 & 369 \\ 
2003-06-11 & NACO/VLT  & 13.24 & Ks        & 71.C-0262(A) & Bailer-Jones & -365.9 &  55.6 & 1533 & 370 \\
2003-06-17 & NACO/VLT  & 13.24 & J, Ks, NB2.17 & 70.C-0476(A) & Bailer-Jones & -367.2 &  54.4 & 1534 & 372 \\
2003-06-27 & NACO/VLT  & 13.24 & J, Ks, NB2.17 & 70.C-0476(A) & Bailer-Jones & -369.8 &  52.5 & 1534 & 375 \\
2003-09-05 & NACO/VLT  & 13.24 & Ks        & 71.C-0262(B) & Bailer-Jones & -386.0 &  37.3 & 1533 & 378 \\
2003-09-14 & NACO/VLT  & 13.24 & J, Ks, NB2.17 & 70.C-0476(A) & Bailer-Jones & -388.3 &  35.1 & 1533 & 381 \\
2003-12-10 & NACO/VLT  & 13.24 & NB1.64        & 72.C-0570(A) & Beuzit       & -405.8 &  14.8 & 1532 & 390 \\
2003-12-12 & NACO/VLT  & 13.24 & J, H, Ks  & 72.C-0022(A) & Leinert      & -406.3 &  14.8 & 1531 & 391 \\
2004-12-11 & NACO/VLT  & 13.24 & Ks        & 74.C-0637(A) & K\"ohler     & -443.6 & -73.3 & 1511 & 446 \\
2006-10-30 & NACO/VLT  & 13.24 & NB2.17        & 78.C-0386(A) & Ratzka       & -347.2 &-197.1 & 1458 & 524 \\
\noalign{\smallskip}
\multicolumn{10}{c}{Own observations}\\
\noalign{\smallskip}
2006-10-09 & CRIRES/VLT  &       &   & 60.A-9078(A) & Reiners     &  & & &  \\
2007-07-16 & NACO/VLT    & 13.24 &NB & 79.C-0106(A) & Neuh\"auser &  -260.7 & -223.7 & 1426 & 563 \\
2007-07-25 & CIAO/SUBARU & 10.82 &NB & o07111       & Neuh\"auser &  -258.2 & -224.4 & 1431 & 567 \\
    \noalign{\smallskip}
    \hline
  \end{tabular}
\end{minipage}
\end{center}

\end{table*}

In addition to the astrometric datapoints published in \citet{Leinert01}, we used archived images from the 
adaptive optics (AO) cameras PUEO/CFHT and NACO/VLT covering a time\-span of nearly 6 years, see 
Table~\ref{tab:observations} for details. We also used the first spatially resolved, high-resolution spectrum
of LHS 1070 ABC \citep{Reiners07c} and determined radial velocities (RV) for the relative
motions of LHS 1070 B and C, as well as for the motion of LHS 1070 A around the barycentre of LHS 1070 BC. 
Finally, in July 2007 we observed LHS 1070 in two imaging campaigns with NACO/VLT and CIAO/SUBARU .

All AO imaging data were reduced by combining individual exposures (shift and add) and determining
the positions and fluxes of each object (including their errors) with IDL/starfinder \citep{starfinder},
a program based on empirical PSF fitting. In cases where a high number of frames were recorded, the 
errors were calculated from statistics over the distribution of the single positions 
obtained from the individual frames instead of the formal error computations on a single co-added frame.

{The pixel scale and detector orientation of PUEO/CFHT were determined from reference images of the 
HIPPARCOS double star HIP482 ($\rho$=7.687\arcsec, $\theta$=148.29\degr) obtained during each of the 
respective runs. The pixel scale was stable at 34.9$\pm$0.1 mas/pixel, but the detector orientation was 
changing between the runs by up to $\pm$2.5\degr, and was stable during each of the runs on the 0.05\degr\,level.
For all NACO/VLT observations taken with the S13 camera, a fixed pixel scale of 13.24$\pm$0.05 mas/pixel, 
and a detector orientation of $\theta = 0.0\pm0.1$\degr was adopted.} Our experience from other long-term 
monitoring programmes \citep[see i.e.][]{Neuh07} show that both, the pixel scale and the detector orientation 
of NACO/VLT are very stable, and the accuracy of their absolute value
is limited by the astrometric errors of available HIPPARCOS double stars rather than instrumental 
instabilities. Finally, our CIAO/SUBARU images were calibrated by comparing an HST/WFPC image of the core
region of the globular cluster M15 with images of the very same region in M15 taken during our run with
CIAO/SUBARU.

We note that for the NACO/VLT observations the dominant error source is the absolute pixel scale 
and detector orientation rather than the centroiding error on the targets. The latter can easily 
be below 1~mas, especially when enough frames with good S/N were obtained. However,
the typical accuracy of NACOs S13 pixel scale is about $\pm$0.05~mas/pix, which in turn limits 
the accuracy of the determined positions to about $\pm$1.5 mas for the typical separation of LHS1070 BC 
($\sim$ 0.5\arcsec) and $\pm$6 mas for the much larger separation of LHS1070 BC around component A. 
Datapoints being closer in time than a few days were combined to a mean epoch since the expected motions 
are far smaller than the derived errors.

The data reduction and calibration of our CRIRES/VLT spectrum is outlined in \citet{Reiners07c}. 
Since here we derive relative radial velocities, our result critically depends on a good dispersion
solution. The covered spectral range falls into a clear atmospheric window. CRIRES is equipped with 
a Th-Ar hollow cathode lamp for wavelength calibration \citep{Kerber07}. More than 2000 lines have been
established as wavelength calibration standards by laboratory measurements in a collaboration between 
the US National Institute of Standards and Technology (NIST) and ESO (Kerber et al. in preparation).
These calibration reference data are implemented in the CRIRES data reduction package. We made use of this 
linelist to calibrate our data.

For the respective relative radial velocity between each of the components of LHS1070,
the alignment of the components in the slit is crucial. At the time of the observation,
all three components are nearly, but not perfectly, aligned along the slit. The misalignment 
(as a deviation from a perfect linear configuration) was $\sim$9\degr. From geometrical considerations (and
by taking the individual S/N ratios into account), we adopt $\pm 1.5$ km/s for the uncertainty in the relative 
RV between LHS 1070 B and C, and for the relative RV between LHS 1070 A and the barycentre of LHS1070 BC. 
We note that the most important constraint for the orbital solution is given by the sign of the radial velocity 
rather than by its exact value.

\section{Analysis and results} 

\subsection{Astrometric solution and orbital fitting: LHS 1070 BC}

\begin{figure}
  \centering
  \resizebox{\hsize}{!}{\includegraphics[clip]{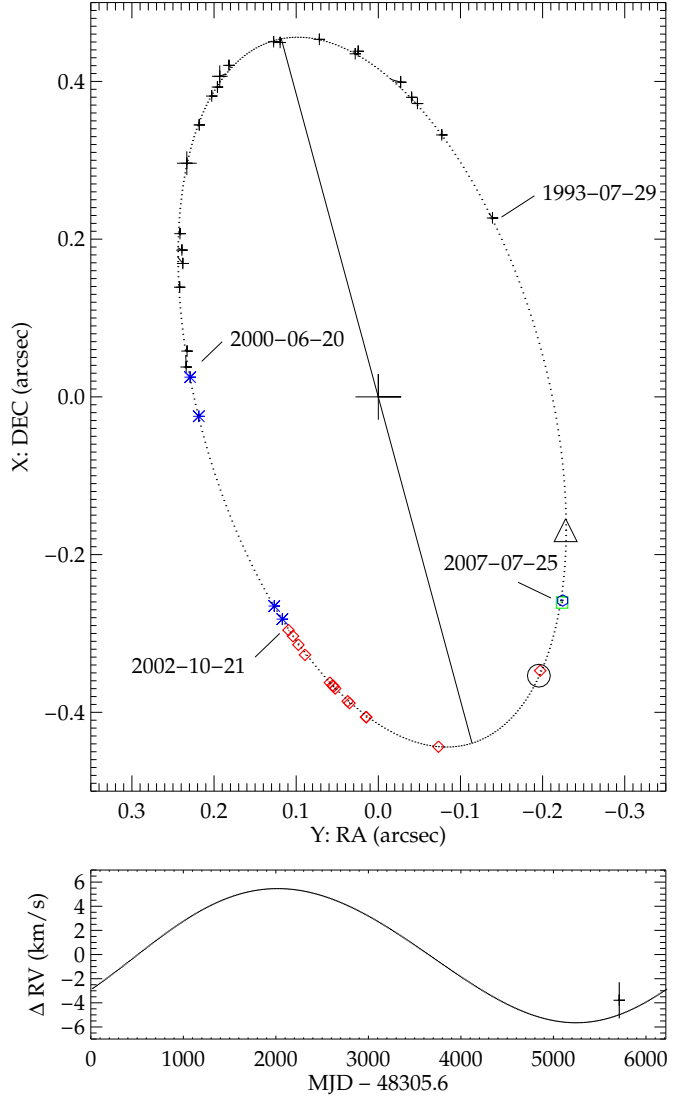}}
  \caption{\label{fig:LHS1070BC_orbit}Relative orbit of LHS~1070 C
around component B. \textit{Upper panel}: astrometric observations from 
\citet{Leinert01} (black crosses) including error bars, PUEO/CFHT 
archival datapoints (blue asterisks), NACO/VLT archival datapoint 
(red diamonds), along with a new NACO/VLT datapoint (green box) 
and a new CIAO/SUBARU datapoint (blue hexagon). Errors and derivations
(O-C) for all datapoints except the original points from 
\citet{Leinert01} are smaller than their respective symbols. 
The time of the radial velocity measurement (2006-10-09) is marked 
with a circle, the periastron with a triangle. The line of nodes is shown. 
\textit{Lower panel}: Relative radial velocity between component C and B. 
The CRIRES/VLT datapoint is shown with its 1$\sigma$ error bar.}
\end{figure}

To fit the relative orbit of LHS1070 B and C around their common barycentre, we chose a similar approach 
to the one presented in \citet{Leinert01}. We simultaneously solved the Keplerian equation using an IDL routine based on 
\citet{Hilditch} for all astrometric and spectroscopic datapoints. A $\chi^2$ minimisation 
was performed to find the optimal set of parameters, and a Jacknife approach \citep{Jacknife} was chosen to 
determine the errors in the orbital parameters (see Table~\ref{tab:small_orbit} for the parameters and
Figure~\ref{fig:LHS1070BC_orbit} for the orbital fit). The orbital coverage is now 253\degr. The 
reduced $\chi^2\simeq1.27$ is reasonable but points towards a slight underestimation of the real
uncertainties in the original dataset of \citet{Leinert01} and in the NACO/VLT data. 

We note that the refined orbital parameters of the BC system agree with the ones 
published in \citet{Leinert01}, but only the time of periastron ($T_0$) {differs slightly, being 
1991-02-18 in our calculation.} The error on the orbital period and semimajor axis is now at the 
sub-percent level, allowing a very precise mass determination for a fixed-distance estimate to the 
system.

\begin{table}[ht]
\begin{minipage}[t]{\columnwidth}
\caption{\label{tab:small_orbit} Orbital and system parameters for LHS~1070\, B and C.}
\centering
\renewcommand{\footnoterule}{}  
\begin{tabular}{lcl}
    \hline
    \hline
    \noalign{\smallskip}
    Parameter & Value & Error \\
    \noalign{\smallskip}
    \hline
    \noalign{\smallskip}  
$a$ (\arcsec)      & 0.4619 &  $\pm$0.0007\\
$e$                & 0.034  &  $\pm$0.002\\
$i$ (\degr)        & 62.59  &  $\pm$0.12\\
$\Omega$ (\degr)   & 14.50  &  $\pm$0.12\\ 
$\omega$ (\degr)   & 240.33 &  $\pm$0.24\\
$P$ (days)         & 6214.74&  $\pm$0.42\\
$T_0$ (MJD)        & 48305.6&  $\pm$1.8\\
&&\\
$\pi_{trig}$ (\arcsec)& 129.47\footnote{from \citet{Costa05}} & $\pm$2.48 \\
$d$ (pc) & 7.72$^{\,a}$ & $\pm$0.15 \\
$a$ (AU) & 3.57 & $\pm$0.07 \\
$M_B+M_C\,(M_{\sun})$ & 0.157 & $\pm$0.009\footnote{including error in distance, at fixed distance the error is 0.0007 $\mathrm{M_{\sun}}$} \\
    \noalign{\smallskip}
    \hline
  \end{tabular}
\end{minipage}
\end{table}

\subsection{Astrometric solution and orbital fitting: LHS 1070 A}

The datapoints for the orbit of the BC system around LHS1070 A presented in \citet{Leinert01}
did not show any strong curvature, hence, the authors used a general $\chi^2$ fit with 
combinations of period and eccentricity to map the parameter space. They used a stability
argument for hierarchical triple systems to further constrain the period and inclination angle.
\begin{figure}
  \centering
  \resizebox{\hsize}{!}{\includegraphics[clip]{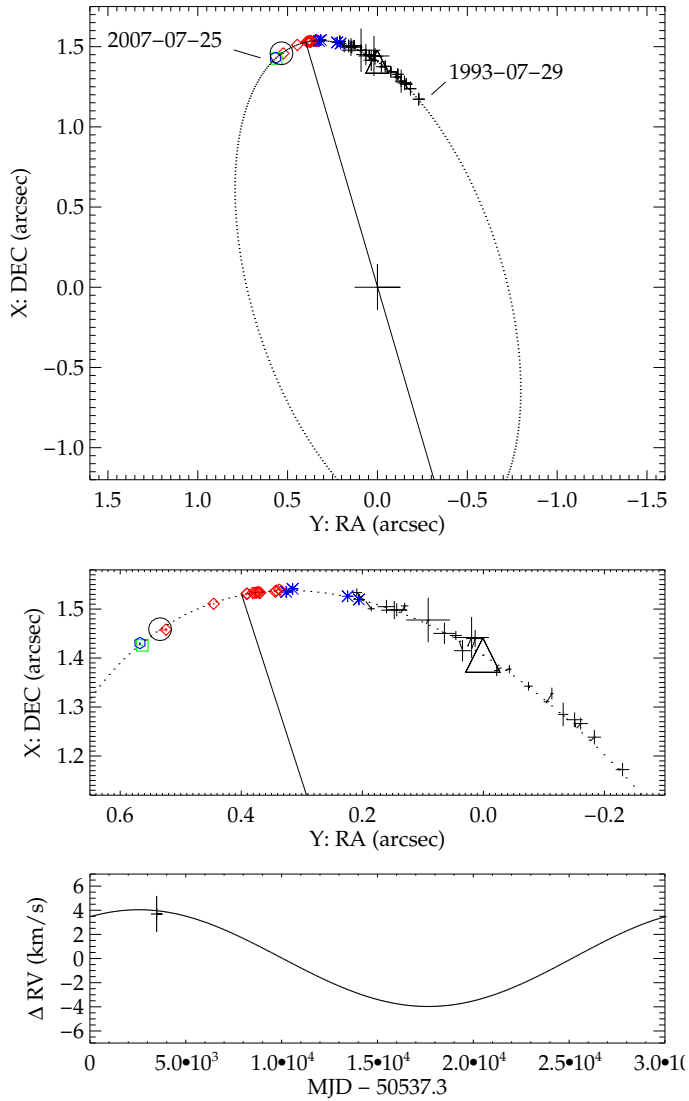}}
  \caption{\label{fig:LHS1070ABC_orbit}Relative orbit of the barycentre of
LHS~1070 B and C around component A. \textit{Upper panel}: astrometric observations.
Same notation and symbols as in Figure~\ref{fig:LHS1070BC_orbit}. 
\textit{Central panel}: Cutout of the upper panel for better visibility.
\textit{Lowest panel}: Relative radial velocity between the barycentre
of LHS~1070 B and C and component A. The CRIRES/VLT datapoint is shown 
with its 1$\sigma$ error bar.}
\end{figure}
The new datapoints reveal a strong curvature, the ascending node being probably close in time 
to our CRIRES/VLT radial velocity datapoint. Thus, the new datapoints give a stronger
constraint on the outer orbit, event hough the covered arc of the orbit is still only about 
33\degr. However, the orbital solution is still not unique. We performed the $\chi^2$ minimisation
for fixed pairs of eccentricity and orbital period, spanning a grid of $e = 0...0.45$ and
$P = 40...120$~yr. Three families of solutions appeared to fit the data: first a small region 
where the eccentricity is zero and the orbital period is $\sim$ 82 years, and second, two narrow
regions for rising eccentricity with either shorter or longer periods with the longitude of periastron
($\omega$) turning by 180\degr\, from $\sim150\degr$ to $\sim327\degr$ between these two 
solutions (see Fig.~\ref{fig:LHS1070A_contour}).

\begin{figure}
  \centering
  \resizebox{\hsize}{!}{\includegraphics[clip]{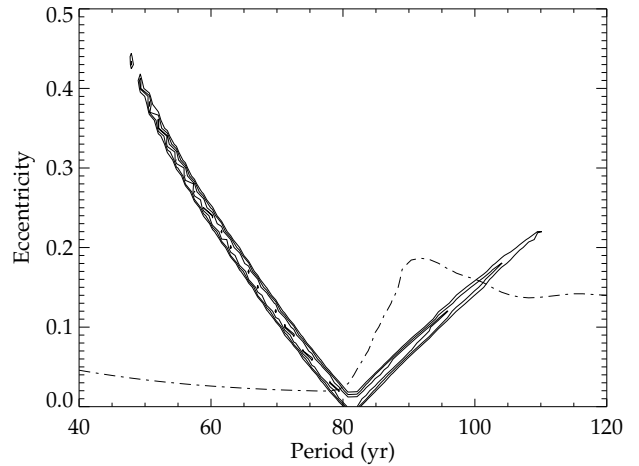}}
  \caption{\label{fig:LHS1070A_contour} The most likely values of eccentricity and orbital 
period for the outer orbit (barycentre of LHS~1070 BC around component A). The contours
show the 1,2, and 3$\sigma$ contours around the $\chi^2$ minimum. The left branch represents 
a family of solutions where $\omega\simeq150\degr$, while the right branch is for a family of 
solutions where $\omega\simeq327\degr$. The dash-dotted line marks the stability threshold of 
the system (see text for details). Only solutions below this line are long-term stable.}
\end{figure}

We have to use a stability criterion to further constrain the parameter
space, as in \citet{Leinert01}. Here we use the critical distance parameter $q_2/a_1$ from 
\citet{Harrington75}, where
$q_2 = a_2(1-e_2)$, with updated values from \citet{Donnison95} to infer the stability limit of 
the LHS 1070 system. Given a prograde orbital motion, the mass estimates of the three bodies and 
solutions for the semimajor axis of the BC system ($a_1$), as well as the values for the eccentricity 
($e_2$) and semimajor axis ($a_2$) of the wide orbit from our grid solution, we can plot the line of 
stability\footnote{Stability is here defined as no significant change in the orbital parameters over
a significant number of orbital revolutions.} 
($q_2/a_1 \geq 3.3$) in the $P$--$e$ plane of Fig.~\ref{fig:LHS1070A_contour}. Only solutions below 
and to the right of this line are long-term stable, excluding most of the regions with orbital
periods shorter than 80 years and eccentricities greater than zero. This is similar to what 
\citet{Leinert01} propose, based on slightly different assumptions.

\begin{figure}
  \centering
  \resizebox{\hsize}{!}{\includegraphics[clip]{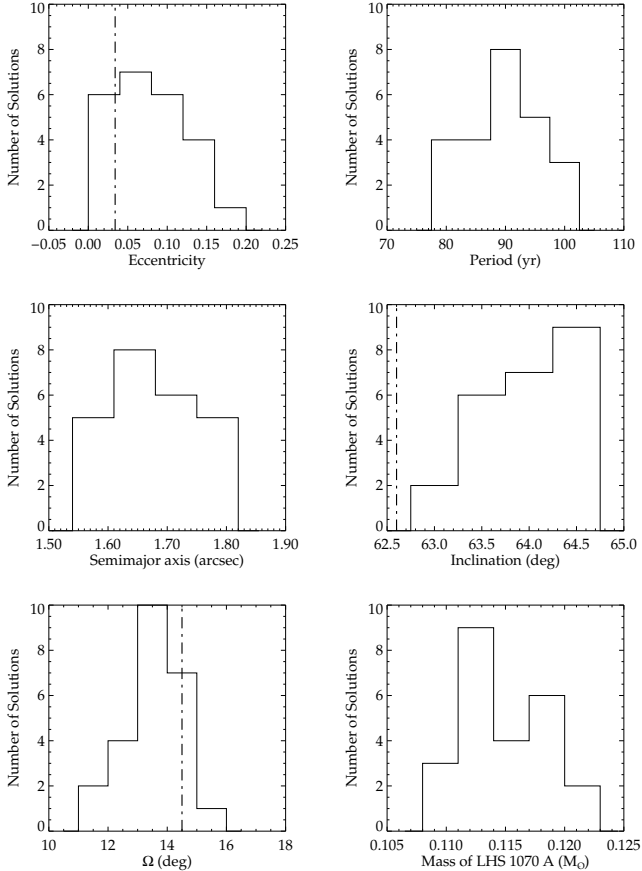}}
  \caption{\label{fig:LHS1070A_contour2} Histograms of the distribution of
orbital parameters from our grid of possible solutions for the outer orbit 
within 2$\sigma$ around min($\chi^2$) and within the stability region 
inferred for the outer orbit. The mass of LHS 1070 A was derived from
the combined system mass and the mass of LHS 1070 BC (see Table 1). The 
spread in mass does not include errors from the distance of the system.
The dashed-dotted lines indicate the respective values from the inner orbit
(LHS 1070 BC) for comparison.}
\end{figure}

Having narrowed down the phase-space of possible solutions for the outer orbit considerably, we
evaluate the histograms of the most important orbital parameters for solutions within 2$\sigma$ around
min($\chi^2$) and within the stability region (see Fig.~\ref{fig:LHS1070A_contour2}). {We find that the 
distribution of the inclination angle for the outer orbit is within 2\degr\, of the value found for the BC orbit ($\sim62.6\degr$) and $\Omega$ is nearly 
identical for both orbits ($\sim14\degr$). Hence, the inner and outer orbits 
appear to be co-planar.} In Fig.~\ref{fig:LHS1070ABC_orbit}
we show the datapoints and the most likely orbital solution with $P\simeq82$~yr and $e$=0.01.

\subsection{Masses}
\begin{figure}
  \centering
  \resizebox{\hsize}{!}{\includegraphics[clip]{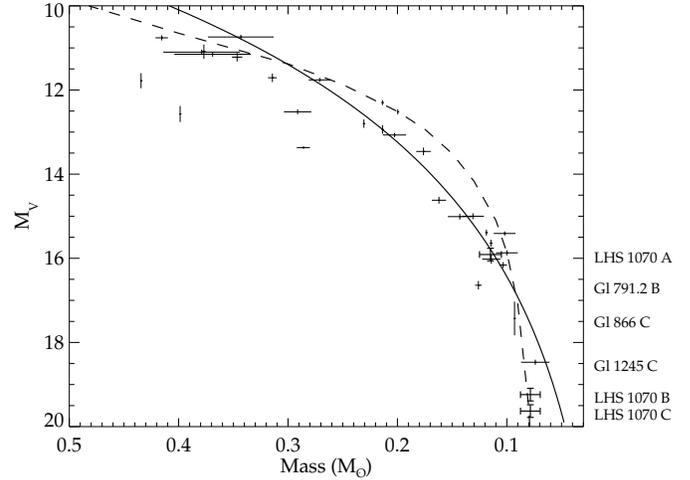}}
  \caption{\label{fig:LHS1070masses}Mass--luminosity relationship at the end of the main sequence. Data taken from
\citet{Delfosse00}, except for GJ 1245 C \citep{Henry99} and LHS 1070 ABC \citep[this paper, $M_V$ from][]{Leinert00}.
The solid line shows the empirical mass--luminosity relationship from \citet[][Equ.~7]{Henry99} and the dashed
line the theoretical relationship from \citet{Baraffe98}.}
\end{figure}
With the new orbit and refined distance \citep{Costa05} of the LHS1070 system, we can
constrain the combined dynamical mass of the B and C components to $M_B+M_C = 0.157 \pm 0.009 
\mathrm{M_{\sun}}$. The dominating error source for the dynamical mass of LHS 1070 BC is still 
the uncertainty of $\pm$0.15~pc in the distance of the system. At a fixed distance of 7.72~pc, 
the error in mass is as small as 0.0007 $\mathrm{M_{\sun}}$. Our value for the combined mass of 
LHS 1070 BC is higher by 9\% than the one obtained by \citet{Leinert01}, mainly because the new 
distance estimate by \citet{Costa05} is higher by nearly 3\% than the value adopted by \citet{Leinert01}. 
This improvement moves the datapoints for LHS 1070 B and C closer to the 5~Gyr isochrone 
in \citet[][Fig.~5]{Leinert01}{; see also Fig.~\ref{fig:LHS1070masses}} here. The combined dynamical mass 
of LHS1070 B and C is thus in very good agreement with the model predictions for the combined mass 
of LHS1070 B and C of 0.159...0.161 $\mathrm{M_{\sun}}$ from \citet{Leinert00} \citep[based on][]{Baraffe98}. 
Hence, both components are slightly above the hydrogen-burning mass limit.

We note that the mass predictions from empirical mass--luminosity relations 
differs from our value by a considerable amount. \citet{Leinert00} give a combined mass for 
LHS1070 B and C of $\simeq 0.138 \mathrm{M_{\sun}}$ from optical, and $\simeq 0.115...0.130 \mathrm{M_{\sun}}$ 
from the near infrared mass--luminosity relations based on \citet{Henry99} and \citet{McCarthy93}.
Such predictions underestimate the true mass by 12 to 26\%. 

The first constraint on the orbit of LHS1070 A around components BC allows the calculation
of a combined mass for the whole system. The error budget for the combined mass of LHS1070 ABC 
is still dominated by the uncertainties in the parameters of the outer orbit (semimajor axis and 
orbital period). However, orbital period and semimajor axis are strongly correlated, and their 
respective error contributions do not simply add up. The combined dynamical mass of the whole 
system of LHS 1070 ABC is $M_{\mathrm{LHS1070}} = 0.272 \pm 0.017 \mathrm{M_{\sun}}$. 

Subtracting the well-constrained dynamical mass of LHS1070 BC from the total dynamical mass of the 
system leaves $M_A = 0.115 \pm 0.010 \mathrm{M_{\sun}}$ for the A component. The precision is
higher than for the total mass estimate of LHS 1070 ABC since the error contribution from the 
uncertainty in distance nearly cancels out. 
The resulting dynamical mass for LHS1070 A is only slightly higher than the predictions, both
from theoretical models of 0.098...0.109~$\mathrm{M_{\sun}}$ and the empirical tracks 
of 0.097...0.113~$\mathrm{M_{\sun}}$ \citep{Leinert00, Baraffe03}.

{In Fig.~\ref{fig:LHS1070masses}
we show the mass--luminosity relation with datapoints from \citet{Delfosse00} and \citet{Henry99}.
LHS 1070 ABC is shown in this graph with absolute $V$ magnitudes obtained from \citet{Leinert00}
and our new mass estimates. Note that the error margin for the mass of LHS 1070 B and C are shown
twice as large to accommodate the uncertainty in the mass ratio of both components. As becomes
apparent from this plot, the empirical mass--luminosity relation is too flat, underpredicting
masses below 0.1~$\mathrm{M_{\sun}}$, while the model prediction from \citet{Baraffe98} is much
steeper and fits the available datapoints.}

\subsection{Is there a fourth component in the system?}
\citet{Henry99} report the discovery of a fourth component in the LHS 1070 system from
HST FGS measurements. The authors give a magnitude difference of $\Delta V \simeq 2.5$
to LHS 1070 A, a mass of 0.07~$M_{\sun}$ for LHS 1070 D, and a separation of $\sim$
50 mas from LHS1070 A at their HST measurements. \citet{Leinert00} find no clear presence 
of such a rather massive, but still much cooler and unresolved component in the spectrum of LHS1070 A. 

We checked all images, especially the ones observed with NACO/VLT under good atmospheric
conditions (hence with a good strehl ratio) for any additional object close to LHS 1070~A. We 
did not find any object at distances equal to or greater than the diffraction limit of the VLT at J, H, 
and Ks band (32 to 56~mas, respectively). We did not find any indication of an elongation or any
difference in the PSF shape of LHS 1070 A, B, and C. This rules out the existence
of a component with more than 0.060~$M_{\sun}$ at distances over 32 mas of 
$\Delta J\leq4.5\mathrm{mag}$ and at distances over 56 mas of $\Delta Ks\leq5.0\mathrm{mag}$, 
based on \citet{Baraffe03} and our detection limits. Given the extended time coverage and tight sampling 
of the NACO/VLT datapoints, it is rather unlikely that we have always missed the object because 
the distance of both components was too small to be resolved.

\citet{Leinert01} argue that the reflex motion of an object of intermediate mass between LHS1070 
A and D should induce a reflex motion of the order of 10~mas, which was not detectable with their 
dataset. Analysing the residuals of the later NACO/VLT measurements, we indeed find a scatter of
of l-3\,mas, exhibiting a non-Gaussian distribution. However, given the large separation of the
observations in comparison to the presumed orbital period of the tentative fourth component, we 
are unable to find an orbital solution for these points. We can only argue that the detected derivations 
are considerably smaller than the prediction of \citet{Leinert01}, implying a much lower 
mass than previously anticipated.

Our mass estimate of LHS 1070 A is in good agreement with the empirical and theoretical predictions.
We do not find any indication of a fourth component in the triple system LHS 1070
with mass down to $\sim$10~$M_{jup}$.

\section{Summary}

Using a combination of published and archival data and adding new observations from high
spatial and spectral resolution imaging and spectroscopy, we have considerably improved the orbital
solution for the low-mass triple system LHS 1070. Most noticeably we found that:
\begin{itemize}

\item The dynamical masses for LHS 1070 B and C, both close to the hydrogen burning limit,
are in good agreement with theoretical predictions from the models of \citet{Baraffe98}
but higher by up to 25\% than the empirical near infrared mass--luminosity relationship of \citet{McCarthy93}.
Since only one datapoint (GJ 1245 C) is constraining the mass--luminosity function below $M_V=18$ 
(see Fig.~\ref{fig:LHS1070masses}), the addition of an accurate value for LHS 1070 BC marks an important 
anchorpoint for the mass--luminosity relation at the end of the main sequence.

\item {The orbits of the LHS 1070 BC system and the outer orbit of LHS 1070 A around the barycentre
of LHS 1070 BC appear co-planar: their inclination angle agrees to within 2\degr\, and the respective 
longitudes of the ascending node ($\Omega$) of both orbits are identical within their uncertainties.}

\item The mass estimate of LHS 1070 A matches the predictions from theory and 
from the empirical tracks. Given its error margin and the fact that no companion could be identified in
the well-resolved VLT/NACO images, we find no indication for a fourth component
(LHS 1070 D) in close orbit around LHS 1070 A.
\end{itemize}

 \begin{acknowledgements}
  We would like to thank the CRIRES science verification team for their
  work on CRIRES and for the execution of the observations. We thank the anonymous
  referee for constructive comments. AS and TR acknowledge financial support from the 
  Deutsche Forschungsgemeinschaft under DFG RE 1664/4-1 and DFG NE 515/23-1, respectively. 
  AR acknowledges financial support through an Emmy Noether Fellowship from the Deutsche 
  Forschungsgemeinschaft under DFG RE 1664/4-1. 
 \end{acknowledgements}

\end{document}